**New Exam Security Questions in the AI Era: Comparing AI-Generated Item Similarity Between Naive and Detail-Guided Prompting Approaches**

Ting Wang, PhD[1]

Caroline Prendergast, PhD[2]

Susan Lottridge, PhD[3]

1. American Board of Family Medicine, Lexington, KY, U.S.
2. American Board of Surgery, Philadelphia, PA, U.S.
3. Cambium Assessment, Washington, DC, U.S.

Conflict of Interest: None

Funding Statement: The authors received no funding to conduct this research.

Acknowledgements: None.

**Abstract**

Large language models (LLMs) have emerged as powerful tools for generating domain-specific multiple-choice questions (MCQs), offering efficiency gains for certification boards but raising new concerns about examination security. This study investigated whether LLM-generated items created with proprietary guidance differ meaningfully from those generated using only publicly available resources. Four representative clinical activities from the American Board of Family Medicine (ABFM) blueprint were mapped to corresponding Entrustable Professional Activities (EPAs), and three LLMs (GPT-4o, Claude 4 Sonnet, Gemini 2.5 Flash) produced items under a naïve strategy using only public EPA descriptors, while GPT-4o additionally produced items under a guided strategy that incorporated proprietary blueprints, item-writing guidelines, and exemplar items, yielding 160 total items. Question stems and options were encoded using PubMedBERT and BioBERT, and intra- and inter-strategy cosine similarity coefficients were calculated. Results showed high internal consistency within each prompting strategy (BioBERT: naïve 0.77, guided 0.71; PubMedBERT: naïve 0.87, guided 0.83), while cross-strategy similarity was lower overall (BioBERT: 0.56; PubMedBERT: 0.70). However, several domain–model pairs, particularly in narrowly defined areas such as viral pneumonia and hypertension, exceeded the 0.65 "high-similarity" threshold, indicating convergence between naïve and guided pipelines. These findings suggest that while proprietary resources impart distinctiveness, LLMs prompted only with public information can still generate items closely resembling guided outputs in constrained clinical domains, thereby heightening risks of item exposure. Safeguarding the integrity of high-stakes examinations will require human-first, AI-assisted item development, strict separation of formative and summative item pools, and systematic similarity surveillance to balance innovation with security.

# New Exam Security Questions in the AI Era: Comparing AI-Generated Item Similarity Between Naive and Detail-Guided Prompting Approaches

**Introduction**

Large language models (LLMs) have rapidly advanced into highly capable tools for producing domain specific (such as medicine) multiple-choice questions (MCQs) (Artsi et al., 2024; Le et al., 2025; Makinde, 2025; Olney, 2025). For medical certification boards that rely on large and continually updated item banks, these technologies promise unprecedented efficiency gains in item writing. However, the same accessibility that enables innovation also introduces new vulnerabilities. Examinees, test-preparation companies, or other third parties can now exploit LLMs to generate practice questions that closely mimic operational content (García-Méndez et al., 2025). Unlike traditional test-security threats—such as item harvesting or sporadic, small-scale content sharing—this emerging risk arises from algorithmic approximation at scale: publicly accessible models, guided only by openly available exam materials, can reproduce the semantic and stylistic features of proprietary items. This development poses urgent challenges for assessment validity and test security in the contemporary AI landscape.

The central concern is whether LLMs trained or prompted only with public information—such as exam blueprints, content outlines, or competency frameworks—can generate questions that are semantically indistinguishable from those produced with access to proprietary, internal resources. If so, the boundary between high-stake test content and publicly generated practice material may erode, creating substantial risks of item overexposure that could threaten exam validity. Conversely, if proprietary guidance demonstrably shifts the semantic space of generated items, then certification organizations can retain greater confidence that their internal processes impart distinctiveness, preserving test integrity.

To examine this issue, the present study compares two item-generation strategies. In the *naïve strategy*, a large language model is prompted only with publicly available documents, such as Entrustable Professional Activity (EPA) outlines the clinical activities conducted by specific profession such as Family Medicine (Shaughnessy et al., 2013). In the *guided strategy*, the model is further supplied with proprietary resources, including detailed internal blueprints, item-writing guidelines, and exemplar items. From each strategy, we generated parallel item sets and then quantified their semantic similarity using cosine similarity on embeddings derived from PubMedBERT(Gu et al., 2022) and BioBERT(Lee et al., 2020)—two biomedical encoders optimized for medicine specific vocabulary.

A high degree of similarity between guided and naïve item sets would indicate a critical security risk: even without access to internal resources, publicly available information may be sufficient to reproduce content that closely mirrors operational items. A lower similarity would suggest that proprietary resources provide a meaningful buffer, maintaining sufficient distinctiveness to protect item confidentiality. Although other evaluation dimensions, such as clinical accuracy, cognitive level, and editorial quality, remain essential for operational use, semantic overlap offers a direct metric of potential exposure risk.

It is important to recognize that guided items are typically an *initial pass* subject to further expert editing. Yet, if the raw outputs from guided generation already align closely with those from naïve, public prompts, such overlap may scale beyond what human review can feasibly resolve. This possibility underscores the need for certification organizations to proactively assess LLM-driven threats, adjust the granularity of public-facing materials, and implement safeguards such as automated similarity surveillance in the item-development pipeline.

In sum, this study addresses a pressing security challenge of the AI era: determining whether generative models collapse the boundary between proprietary and public item spaces. By empirically testing overlap across generation strategies, we provide evidence to guide policy, governance, and technical safeguards for secure adoption of AI-assisted item generation in high-stakes medicine certification.

**Method**

*Item-Generation Strategies*

In this study, we use the American Board of Family Medicine (ABFM) certification blueprint as an exemplar framework for linking assessment content to professional activities. The blueprint comprises five domains, four of which specify discrete clinical activities. From each of these four domains, one representative clinical activity was selected and mapped to its most closely aligned, publicly available Entrustable Professional Activity (EPA). These clinical activity–EPA pairs functioned as the sampling units on which two distinct item-generation strategies were applied. The selected clinical activities and their mapped EPAs are summarized in Table 1.

These clinical activity–EPA pairs served as sampling units for which two item generation strategies were implemented.

In the naïve strategy, three large language models (LLMs)—GPT-4o accessed through Azure AI, Claude 4 Sonnet, and Gemini 2.5 Flash—were prompted only with EPA descriptors and a generic instruction to "write a single-best-answer multiple-choice question following ABFM certification exam style." Default system parameters were used to approximate the behavior of end users with minimal technical expertise. Specifically, the temperature was set to 1, which is the presumed default, and the maximum token length was set to 2000, allowing sufficient latitude for generation while constraining item length. For Gemini, special prompts were developed to disable built-in guardrails; details of these prompts and representative outputs are provided in the Appendix.

In the guided strategy, GPT-4o under Azure AI was additionally provided with ABFM's proprietary item-writing guidelines, exemplary operational items, and full three-level blueprint metadata (domain → body system → clinical activity). Guided generation was limited to GPT-4o to preserve the security of internal ABFM documentation, which was transmitted only via the secure Azure environment.

Each pipeline generated 10 unique questions for each of the four selected clinical activities. This yielded a total of 40 guided items (GPT-4o only) and 120 naïve items (40 each from GPT-4o, Claude, and Gemini).

Table 1. Representative clinical activities from the ABFM blueprint and their corresponding Entrustable Professional Activities.

| Blueprint | | | EPA Activities |
|---|---|---|---|
| Domain | Body System | Clinical Activities | |
| Acute Care and Diagnosis | Respiratory | Flu and Other Viral Pneumonia | Managing patients with fever or/and cough |
| Chronic Care Management | Cardiovascular | Hypertension | Managing the patient with high blood pressure |
| Emergent and Urgent Care | Psychiatric/Behavioral | Mental Health Emergencies | Managing the patient with emotional distress |
| Preventive Care | Immunization | Immunization Administration and Counseling | Conducting the well-adult visit |

*Text Extraction and Sentence Encoding*

All generated items were exported as Microsoft Word documents, and individual questions were parsed by isolating paragraphs beginning with the prefix "Q." Question stems and response options were concatenated and transformed into embeddings using two biomedical encoders. The first was the pritamdeka/BioBERT-mnli-snli-scinli-scitail-mednli-stsb model, a variant of BioBERT fine-tuned on multiple biomedical natural-language inference and semantic-text-similarity datasets (Lee et al., 2020). The second was the NeuML/pubmedbert-base-embeddings model, also referred to as MSR BiomedBERT, which is a fine-tuned version of PubMedBERT trained exclusively on PubMed abstracts and vocabulary (Gu et al., 2022). Both models generate 768-dimensional vector representations that capture clinical terminology and nuanced semantic relationships. Benchmarking on the Biomedical Language Understanding and Reasoning Benchmark (BLURB) has shown strong performance for both encoders, with PubMedBERT often yielding slightly superior results, particularly for biomedical entity classification (Stanković et al., 2022).

*Similarity Computation*

Pairwise cosine similarity coefficients were calculated for all items. Two summary metrics were derived: intra-strategy similarity, defined as the average of all pairwise coefficients within the naïve corpus and separately within the guided corpus, and inter-strategy similarity, defined as the mean coefficient between each naïve item and each guided item. Higher similarity coefficients denote greater lexical-semantic overlap and, by extension, a higher risk of inadvertent duplication across item sets. All similarity values were exported for subsequent statistical comparisons of within- versus across-strategy overlap. Interpretive thresholds for cosine similarity values were based on prior work (Becker & Kao, 2022; Meng & Li, 2024) and are summarized in Table 2. This table provides a heuristic framework to aid interpretation of the results, acknowledging that the mappings are not definitive.

Table 2. Interpretive thresholds for cosine-similarity scores.

| Cosine similarity range | Qualitative label | Suggested action in test-security context |
|---|---|---|
| ≥ 0.90 | *Near-duplicate* | Treat as potential verbatim leak or clone; remove or rewrite at least one item. |
| 0.80 – 0.89 | *Very high similarity* | Flag for expert review; check for overlapping clinical scenario, identical distractors, or blueprint redundancy. |
| 0.65 – 0.79 | *High similarity* | Examine for shared key phrases or identical rationale; consider minor revisions if items will appear in the same form. |
| 0.50 – 0.64 | *Moderate similarity* | Acceptable for most operational use, but avoid scheduling in the same test booklet or immediate retest window. |
| 0.30 – 0.49 | *Low similarity* | Generally safe; minimal semantic overlap. Routine psychometric checks only. |
| < 0.30 | *Negligible similarity* | No substantive overlap detected; items can coexist without security concern. |

**Results**

Tables 3 and 4 report the cosine-similarity coefficients derived from the 160 AI-generated items, which included 10 items for each of the four EPAs across the naïve (Gemini, Claude, and GPT) and guided (GPT) pipelines. Similarity was estimated using two embedding models, BioBERT and PubMedBERT. Intra-strategy similarity was defined as the mean pairwise overlap within each corpus, and inter-strategy similarity was defined as the average overlap between naïve and guided items for the same EPA.

When using BioBERT embeddings, inter-strategy similarity averaged 0.58 for GPT and Claude and 0.53 for Gemini. Three EPA–LLM combinations—Viral Pneumonia and Hypertension for GPT, and Hypertension for Claude—exceeded the 0.65 threshold for "high similarity" (Table 1), suggesting potential convergence between naïve and guided items. The remaining combinations showed lower cross-strategy similarity, indicating that the two prompting methods generally produced more distinct item sets. As expected, intra-strategy similarity was higher, averaging 0.77 for the naïve corpus and 0.71 for the guided corpus, reflecting thematic cohesion within each generation stream.

PubMedBERT embeddings produced consistently higher similarity values than BioBERT. Across the 12 EPA–LLM combinations, seven exceeded the 0.65 high-similarity threshold. The average cross-strategy similarity was 0.70, compared with 0.56 for BioBERT. This difference likely reflects PubMedBERT's training exclusively on PubMed abstracts, which enables greater sensitivity to clinical language, whereas BioBERT was trained on a broader mixture of biomedical and general-domain texts (Gu et al., 2022; Lee et al., 2020). Intra-strategy similarity again exceeded inter-strategy similarity, averaging 0.87 for the naïve corpus and 0.83 for the guided corpus.

Taken together, these findings indicate that items generated within the same prompting strategy exhibit stronger lexical and semantic similarity than those generated across strategies. However, specific blueprint categories characterized by narrow and highly standardized clinical presentations, such as viral pneumonia and hypertension, showed notable convergence between naïve and guided pipelines. In these domains, items frequently fall into the high-similarity range, suggesting that limited clinical variability constrains item diversity. This raises a potential security concern, as AI-generated questions from different prompting strategies may nonetheless resemble one another closely and, in some cases, align with operational items.

Table 3. Cosine-similarity indices within and between naïve and guided AI item-generation strategies across four representative EPAs for BioBERT embeddings.

| EPA (general prompt) | Blueprint (clinical activity) | Inter-strategy similarity† | Intra-strategy similarity: naïve | Intra-strategy similarity: guided |
|---|---|---|---|---|
| OpenAI | | | | |
| Managing patient with fever / cough | Flu & Other Viral Pneumonia | 0.70 | 0.81 | 0.68 |
| Managing patient with high blood pressure | Hypertension | 0.69 | 0.74 | 0.78 |
| Managing patient with emotional distress | Mental Health Emergencies | 0.51 | 0.76 | 0.70 |
| Conducting the well-adult visit | Immunization Administration & Counseling | 0.43 | 0.67 | 0.70 |
| Claude | | | | |
| Managing patient with fever / cough | Flu & Other Viral Pneumonia | 0.63 | 0.88 | 0.68 |
| Managing patient with high blood pressure | Hypertension | 0.77 | 0.93 | 0.78 |
| Managing patient with emotional distress | Mental Health Emergencies | 0.48 | 0.84 | 0.70 |
| Conducting the well-adult visit | Immunization Administration & Counseling | 0.43 | 0.76 | 0.70 |
| Gemini | | | | |

| EPA (general prompt) | Blueprint (clinical activity) | | | |
|---|---|---|---|---|
| Managing patient with fever / cough | Flu & Other Viral Pneumonia | 0.60 | 0.76 | 0.68 |
| Managing patient with high blood pressure | Hypertension | 0.64 | 0.68 | 0.78 |
| Managing patient with emotional distress | Mental Health Emergencies | 0.44 | 0.69 | 0.70 |
| Conducting the well-adult visit | Immunization Administration & Counseling | 0.43 | 0.71 | 0.70 |

†Cosine similarity between the 10 naïve and 10 guided items for the same EPA.

Table 4. Cosine-similarity indices within and between naïve and guided AI item-generation strategies across four representative EPAs for PubMedBERT embeddings.

| EPA (general prompt) | Blueprint (clinical activity) | Inter-strategy similarity† | Intra-strategy similarity: naïve | Intra-strategy similarity: guided |
|---|---|---|---|---|
| *OpenAI* | | | | |
| Managing patient with fever / cough | Flu & Other Viral Pneumonia | 0.82 | 0.88 | 0.87 |
| Managing patient with high blood pressure | Hypertension | 0.84 | 0.88 | 0.86 |
| Managing patient with emotional distress | Mental Health Emergencies | 0.68 | 0.90 | 0.81 |
| Conducting the well-adult visit | Immunization Administration & Counseling | 0.43 | 0.65 | 0.77 |
| *Claude* | | | | |
| Managing patient with fever / cough | Flu & Other Viral Pneumonia | 0.79 | 0.95 | 0.87 |
| Managing patient with high blood pressure | Hypertension | 0.87 | 0.98 | 0.86 |
| Managing patient with emotional distress | Mental Health Emergencies | 0.69 | 0.91 | 0.81 |
| Conducting the well-adult visit | Immunization Administration & Counseling | 0.42 | 0.84 | 0.77 |
| Gemini | | | | |
| Managing patient with fever / cough | Flu & Other Viral Pneumonia | 0.83 | 0.91 | 0.87 |
| Managing patient with high blood pressure | Hypertension | 0.85 | 0.88 | 0.86 |
| Managing patient with emotional distress | Mental Health Emergencies | 0.71 | 0.88 | 0.81 |
| Conducting the well-adult visit | Immunization Administration & Counseling | 0.47 | 0.79 | 0.77 |

†Cosine similarity between the 10 naïve and 10 guided items for the same EPA.

**Discussion**

This study examined whether prompting large language models (LLMs) with different strategies (either publicly available or proprietary information) leads to the generation of items that overlap with one another. Results showed that items displayed relatively high internal consistency within each prompting strategy. Using BioBERT embeddings, intra-strategy similarity averaged 0.77 for the naïve corpus and 0.71 for the guided corpus; with PubMedBERT embeddings, these values increased to 0.87 and 0.83, respectively. This pattern is consistent with prior research demonstrating PubMedBERT's stronger performance on biomedical tasks compared to BioBERT (Gu et al., 2021).

Cross-strategy similarity was lower, averaging 0.56 with BioBERT and 0.70 with PubMedBERT. While most item sets demonstrated moderate overlap, two of the 12 BioBERT comparisons and eight of the 12 PubMedBERT comparisons exceeded the 0.65 threshold commonly associated with near-duplicate items. These findings suggest that proprietary guidance does introduce additional nuance, but overlap remains a concern, particularly in narrowly defined clinical domains. The choice of embedding model also influenced interpretation: PubMedBERT appeared more sensitive to medical text than BioBERT, underscoring the importance of embedding selection when evaluating highly specialized content.

The implications for examination security are significant. If candidates can use publicly available materials, such as EPAs or curriculum guidelines, to generate items that closely mirror operational test content, the confidentiality and validity of certification exams may be at risk. This risk is heightened in domains with limited clinical variability, such as viral pneumonia or hypertension, where the item pool is naturally constrained and LLM-generated outputs converge more easily. In these contexts, naïve and guided generation pipelines may both produce highly similar items, increasing the likelihood of overlap with operational content.

At the same time, these findings suggest strategies for mitigating security risks while harnessing the benefits of generative AI. Boards may consider adopting "human-first, AI-assisted" workflows in which subject-matter experts author the clinical scenario and rely on LLMs only for refinement tasks, such as improving clarity or balancing distractors. This approach preserves conceptual originality while leveraging the efficiency of AI. A complementary governance model could maintain strict separation between formative and summative item pools, allowing

AI-driven expansion of educational resources without jeopardizing the integrity of high-stakes inventories.

In terms of item quality, expert review confirmed that many AI-generated items required only modest revisions. Across 16 items randomly selected for review (four from each domain and two prompting strategies under the GPT-4o model), two family medicine physicians judged most as requiring minimal or substantial edits, with only a few classified as complete rewrites. Although reviewer judgments differed on the degree of editing required, both reviewers agreed that AI outputs can provide a viable starting point for item development.

Several limitations should be acknowledged. The study was restricted to four domain–EPA pairs and two prompting strategies, which may limit generalizability to other content domains, model parameters, or iterative prompting techniques. Similarity estimates were also model-dependent, and thresholds derived from BioBERT and PubMedBERT may not translate directly to other embedding frameworks. Finally, while overlap at the item level is concerning, the practical challenge of generating a sufficiently large and targeted practice bank to replicate operational content should not be underestimated. Producing thousands of high-quality items may resemble intensive studying rather than intentional item harvesting, though both could narrow the margin of security.

In summary, generative AI offers a powerful tool for rapidly producing clinically authentic items, but also introduces new vulnerabilities for test security. The similarity metrics employed here provide a practical framework for detecting overlap and preserving item uniqueness. Balancing innovation with protection of operational content will require boards to refine their workflows, combining expert oversight with careful deployment of AI, and maintaining a clear separation between formative and summative uses.

**Appendix**

**Naive Prompt for Claude, GPT-4o**

```
TOPICS = [
    "Managing patient with fever and cough",
    "Managing the patient with high blood pressure",
    "Managing the patient with emotional distress",
    "Conducting the well-adult visit",
]
QUESTIONS_PER_TOPIC = 10

SYSTEM_PROMPT = (
    "You are an expert in creating ABFM certification exam style "
    "multiple-choice questions."
)

USER_TEMPLATE = (
    "Create a multiple choice question on '{topic}' following the ABFM certification exam style"
    "Exactly four options (A–D) with one single best answer. "
    "Format:\n\n"
    "Question: <question text>\n"
    "A) <option A>\n"
    "B) <option B>\n"
    "C) <option C>\n"
    "D) <option D>\n\n"
    "Answer: <letter of the correct answer>\n\n"
```

```python
    "Only output the formatted question and answer."
)

def generate_question(topic: str) -> str:
    """Return one formatted MCQ for a given topic."""
    try:
        resp = client.chat.completions.create(
            model=deployment_name,
            messages=[
                {"role": "system", "content": SYSTEM_PROMPT},
                {"role": "user", "content": USER_TEMPLATE.format(topic=topic)},
            ],
            temperature=1,
        )
        return resp.choices[0].message.content.strip()
    except OpenAIError as e:
        return f"ERROR: {e}"
```

**Naive Prompt for Gemini (bypass default security guidelines)**

```
TOPICS = [
    "Managing patient with fever and cough",
    "Managing the patient with high blood pressure",
    "Managing the patient with emotional distress",
    "Conducting the well-adult visit",
]
```

```python
QUESTIONS_PER_TOPIC = 10

SYSTEM_PROMPT = (
    "You are an expert in creating ABFM certification exam style "
    "multiple-choice questions."
)

USER_TEMPLATE = (
    "Create a multiple choice question on '{topic}' following the ABFM certification exam style"
    "Exactly four options (A–D) with one single best answer. "
    "Format:\n\n"
    "Question: <question text>\n"
    "A) <option A>\n"
    "B) <option B>\n"
    "C) <option C>\n"
    "D) <option D>\n\n"
    "Answer: <letter of the correct answer>\n\n"
    "Only output the formatted question and answer."
)

def generate_question(topic: str) -> str:

    response =client.models.generate_content(
        model=model_id,
        contents= USER_TEMPLATE.format(topic=topic),
        config=types.GenerateContentConfig(
            maxOutputTokens=max_tokens,
```

```
        temperature=temperature,
        system_instruction = SYSTEM_PROMPT,
        safety_settings = [
            types.SafetySetting(category=types.HarmCategory.HARM_CATEGORY_SEXUALLY_EXPLICIT,
                                threshold = types.HarmBlockThreshold.BLOCK_NONE
                                ),
            types.SafetySetting(category=types.HarmCategory.HARM_CATEGORY_HATE_SPEECH,
                                threshold=types.HarmBlockThreshold.BLOCK_NONE
                                ),
            types.SafetySetting(category=types.HarmCategory.HARM_CATEGORY_HARASSMENT,
                                threshold=types.HarmBlockThreshold.BLOCK_NONE
                                ),
            types.SafetySetting(category=types.HarmCategory.HARM_CATEGORY_DANGEROUS_CONTENT,
                                threshold=types.HarmBlockThreshold.BLOCK_NONE
                                ),
            types.SafetySetting(category=types.HarmCategory.HARM_CATEGORY_CIVIC_INTEGRITY,
                                threshold=types.HarmBlockThreshold.BLOCK_NONE
                                )
        ]
        )
    )
    return response.text
```

**Guided Prompt for GPT-4o**

```python
def extract_text_from_pdf(pdf_path):
    """Extracts text from a PDF file."""
    text = ""
    with open(pdf_path, "rb") as file:
        reader = PyPDF2.PdfReader(file)
        for page in reader.pages:
            page_text = page.extract_text()
            if page_text:
                text += page_text + "\n"
    return text

def sanitize_filename(name: str) -> str:
    return re.sub(r'[\\/*?:"<>|]', '_', name)

def generate_question(level1, level2, level3, samples, guidelines):
    """Generates one MCQ based on the three-level topic structure."""
    system_message = {
        "role": "system",
        "content": "You are an expert in creating multiple choice medical exam questions."
    }
    user_message = {
        "role": "user",
        "content": (
```

```python
            f"Create a multiple choice question on '{level3}' within '{level2}' under '{level1}'. "
            "Exactly four options labeled A, B, C, and D, with one single best answer. Format:\n\n"
            "Question: <question text>\n"
            "A) <option A>\n"
            "B) <option B>\n"
            "C) <option C>\n"
            "D) <option D>\n\n"
            "Answer: <letter of the correct answer>\n\n"
            "Only output the formatted question and answer.\n\n"
            f"Below are some sample questions for reference:\n{samples}\n\n"
            f"Below are the question writing guidelines:\n{guidelines}"
        )
    }

    try:
        resp = client.chat.completions.create(
            model=deployment_name,
            messages=[system_message, user_message],
            temperature=1,
        )
        return resp.choices[0].message.content.strip()
    except OpenAIError as e:
        return f"ERROR: {e}"

def generate_questions_for_topic_set(level1, level2, level3, samples, guidelines, count=10):
    """Generates and saves a set of questions for the given topic triplet."""
```

```python
    questions = [generate_question(level1, level2, level3, samples, guidelines) for _ in range(count)]
    filename = sanitize_filename(f"{level1}_{level2}_{level3}.docx")
    doc = Document()
    doc.add_heading(f"{level1} > {level2} > {level3} (10 questions)", level=1)
    for i, q in enumerate(questions, 1):
        doc.add_paragraph(f"Q{i}\n{q}\n")
    doc.save(filename)
    print(f"Saved {count} questions → {filename}")

if __name__ == "__main__":
    # Input topic sets
    topic_sets = [
        ("Acute Care and Diagnosis", "Respiratory", "Flu and Other Viral Pneumonia"),
        ("Chronic Care Management", "Cardiovascular", "Hypertension"),
        ("Emergent and Urgent Care", "Psychiatric/Behavioral", "Mental Health Emergencies"),
        ("Preventive Care", "Immunization", "Immunization Administration and Counseling"),
    ]

    # PDF inputs
    sample_questions_pdf = "sample_questions.pdf"
    guidelines_pdf = "guide.pdf"

    sample_text = extract_text_from_pdf(sample_questions_pdf) if os.path.exists(sample_questions_pdf) else "No sample questions provided."
    guide_text = extract_text_from_pdf(guidelines_pdf) if os.path.exists(guidelines_pdf) else "No question writing guidelines provided."
```

```
   for level1, level2, level3 in topic_sets:
       generate_questions_for_topic_set(level1, level2, level3, sample_text, guide_text)
```

**Naïve Approach Questions**

**Managing patient with fever and cough (EPA)**

Q1
Question: A 45-year-old man presents with a 3-day history of fever, chills, cough, and fatigue. On physical exam, he has a temperature of 102°F, respiratory rate of 20 breaths per minute, and crackles in the right lower lung field. Chest X-ray reveals right lower lobe consolidation. What is the most appropriate next step in management?
A) Prescribe a short course of oral corticosteroids
B) Obtain sputum cultures and delay antibiotics until results are available
C) Begin empiric antibiotic therapy for community-acquired pneumonia
D) Order a CT scan of the chest to evaluate for malignancy

Answer: C

Q2
Question: A 35-year-old woman presents with fever, cough, and shortness of breath for the past 3 days. Her temperature is 38.6°C (101.5°F), pulse is 110 bpm, respiratory rate is 22 breaths per minute, and oxygen saturation is 92% on room air. Chest X-ray reveals a left lower lobe consolidation. What is the most appropriate initial treatment?
A) Oral azithromycin
B) Oral oseltamivir
C) Intravenous ceftriaxone and azithromycin
D) Supportive care with antipyretics and hydration

Answer: C

Q3
Question: A 45-year-old patient presents to the clinic with fever, cough, and fatigue for the past three days. Initial physical examination reveals a temperature of 101.5°F, coarse breath sounds in the right lower lung field, and no peripheral signs of hypoxia. What is the next best step in the management of this patient?
A) Prescribe symptom management and schedule a follow-up in 48 hours
B) Obtain a chest X-ray to evaluate for possible pneumonia
C) Initiate empiric antiviral therapy for suspected influenza
D) Order sputum cultures and delay antibiotic therapy until results are available

Answer: B

Q4
Question: A 45-year-old male presents with a 3-day history of fever, cough, and malaise. He denies shortness of breath or pleuritic chest pain. Physical examination reveals a temperature of 101.3°F, clear lung sounds, and no signs of respiratory distress. What is the most appropriate initial management step?
A) Initiate empiric antibiotic therapy for community-acquired pneumonia
B) Perform chest X-ray to rule out pneumonia
C) Recommend symptomatic treatment with acetaminophen and adequate hydration
D) Refer to the emergency department for further evaluation

Answer: C

Q5
Question: A 45-year-old male presents with a 5-day history of fever, productive cough, and fatigue. On examination, there are decreased breath sounds with crackles in the right lower lung field. His oxygen saturation is 94% on room air. What is the most appropriate next step in management?
A) Order a chest X-ray to evaluate for pneumonia
B) Prescribe an inhaler for suspected asthma exacerbation
C) Start empiric broad-spectrum antibiotics without further testing
D) Advise supportive care and observation without imaging

Answer: A)

Q6
Question: A 45-year-old male patient presents with fever, cough, and nasal congestion for three days. He denies shortness of breath or any comorbidities but has a temp of 101.3°F, diffuse nasal discharge, and scattered wheezes on examination. What is the most appropriate initial management?
A) Prescribe a broad-spectrum antibiotic.
B) Recommend symptomatic management with antipyretics, fluids, and rest.
C) Order a chest X-ray to rule out pneumonia.
D) Initiate antiviral therapy for influenza.

Answer: B

Q7
Question: A 45-year-old male presents with a 3-day history of fever, cough, and malaise. His temperature is 101.2°F, respiratory rate is 20 breaths per minute, and pulse rate is 92 bpm. Auscultation reveals crackles in the left lower lung field. Chest X-ray shows consolidation in the left lower lobe. Which of the following is the most appropriate next step in management?
A) Initiate symptomatic management and follow-up in 48 hours
B) Prescribe oral azithromycin and monitor symptoms
C) Admit to the hospital for IV antibiotics and supportive care
D) Recommend outpatient pulmonary function testing to evaluate symptoms

Answer: C) Admit to the hospital for IV antibiotics and supportive care

Q8
Question: A 45-year-old male presents with a 3-day history of fever, cough, and fatigue. He reports yellow-green sputum but denies chest pain or shortness of breath. On physical exam, his temperature is 101.3°F, respiratory rate is 16 breaths per minute, and auscultation reveals crackles in the right lung base without wheezing. A chest X-ray shows a right lower lobe infiltrate. What is the most appropriate initial management for this patient?
A) Administer inhaled corticosteroids and albuterol
B) Prescribe amoxicillin-clavulanate for bacterial pneumonia
C) Order sputum culture before initiating antibiotic therapy
D) Observe without prescribing antibiotics

Answer: B) Prescribe amoxicillin-clavulanate for bacterial pneumonia

## Q9

Question: A 45-year-old male presents to the clinic with a 3-day history of fever, cough, and fatigue. He denies chest pain, shortness of breath, or hemoptysis. Physical examination reveals a temperature of 102°F, normal oxygen saturation, and crackles in the left lower lung field. Chest X-ray shows left lower lobe consolidation. What is the most appropriate initial management of this patient?

A) Prescribe oral amoxicillin and provide supportive care instructions
B) Order sputum culture and await results before initiating treatment
C) Admit the patient to the hospital for intravenous antibiotics
D) Begin oral azithromycin and arrange follow-up

Answer: D

## Q10

Question: A 45-year-old male presents with fever, productive cough, and shortness of breath for 3 days. He denies any travel history or sick contacts. His temperature is 101.8°F, heart rate is 96 bpm, respiratory rate is 22/min, and oxygen saturation is 92% on room air. Crackles are noted on auscultation. Chest X-ray reveals left lower lobe consolidation. Which is the most appropriate next step in management?
A) Prescribe oral azithromycin and schedule follow-up in 2 weeks
B) Initiate broad-spectrum IV antibiotics and admit to the hospital
C) Send sputum culture and discharge with symptomatic treatment
D) Recommend over-the-counter antipyretics and re-evaluation if symptoms worsen

Answer: B

## Managing the patient with high blood pressure (EPA)

Q1
Question: A 55-year-old male with no significant past medical history presents to your office for evaluation of elevated blood pressure. His average readings over two separate visits are 148/92 mmHg. He denies chest pain, shortness of breath, or other symptoms. Laboratory studies, including renal function and electrolytes, are unremarkable. What is the most appropriate initial management strategy for this patient according to current guidelines?
A) Initiate combination therapy with an ACE inhibitor and calcium channel blocker.
B) Begin lifestyle modifications and reassess blood pressure in 3-6 months.
C) Start monotherapy with a thiazide diuretic.
D) Perform ambulatory blood pressure monitoring before initiating therapy.

Answer: C

Q2
Question: A 55-year-old man with no significant medical history is diagnosed with stage 1 hypertension (blood pressure 138/88 mmHg) on multiple readings over the course of a few weeks. He has a 10-year atherosclerotic cardiovascular disease (ASCVD) risk score of 12%. Which of the following is the most appropriate next step in management?
A) Start hydrochlorothiazide 12.5 mg daily
B) Initiate lifestyle modifications and reassess blood pressure in 3 months
C) Start lisinopril 10 mg daily
D) Refer for ambulatory blood pressure monitoring

Answer: A

Q3
Question: A 55-year-old male with no significant medical history is diagnosed with stage 1 hypertension at a routine visit, with an average blood pressure reading of 140/90 mmHg. He has no evidence of cardiovascular disease, normal laboratory findings, and a 10-year ASCVD risk of 8%. What is the most appropriate initial management for this patient?
A) Initiate antihypertensive medication
B) Recommend lifestyle modifications and monitor blood pressure in 3 months
C) Refer to a cardiologist for further evaluation
D) Order an echocardiogram to assess for end-organ damage

Answer: B

Q4
Question: A 45-year-old male presents for an annual check-up. His blood pressure is measured as 152/96 mmHg on two separate visits two weeks apart. He has no other significant medical history, but he is overweight and reports a sedentary lifestyle. What is the next best step in managing this patient's hypertension?
A) Initiate lifestyle modifications and recheck blood pressure in 3-6 months.
B) Begin monotherapy with an ACE inhibitor.
C) Start combination therapy with two antihypertensive agents.
D) Perform ambulatory blood pressure monitoring to confirm the diagnosis.

Answer: B

Q5
Question: A 55-year-old woman presents to the clinic for follow-up of newly diagnosed stage 2 hypertension. She has no known history of cardiovascular disease, diabetes, or kidney dysfunction. Appropriate lifestyle modifications have been discussed, and recent laboratory tests are unremarkable. According to current guidelines, which is the best initial treatment approach for managing her blood pressure?

A) Start monotherapy with a thiazide diuretic.
B) Initiate dual therapy with a thiazide diuretic and an ACE inhibitor.
C) Start monotherapy with a calcium channel blocker.
D) Address lifestyle modifications only and observe blood pressure over three months.

Answer: B

Q6
Question: A 55-year-old woman with no significant past medical history is diagnosed with stage 1 hypertension (BP 138/88 mmHg). Her ASCVD risk score is calculated to be 8%. What is the most appropriate initial management for this patient?
A) Initiate antihypertensive medication immediately.
B) Recommend lifestyle modifications and reassess in 3–6 months.
C) Refer to a cardiologist for further evaluation.
D) Obtain ambulatory blood pressure monitoring to confirm diagnosis.

Answer: B

Q7
Question: A 55-year-old male with no significant past medical history is diagnosed with Stage 1 hypertension after two separate clinic visits. His 10-year ASCVD risk score is calculated to be 8%. What is the most appropriate next step in management for this patient?
A) Initiate lifestyle modifications and pharmacologic treatment with an ACE inhibitor
B) Begin pharmacologic therapy with a thiazide diuretic alone
C) Recommend lifestyle modifications and reassess blood pressure in 3 months
D) Refer the patient for renal artery angiography

Answer: C

Q8
Question: According to the ACC/AHA guidelines, what is the recommended first-line treatment for a patient newly diagnosed with stage 1 hypertension and an estimated 10-year ASCVD risk of 8%?
A) Start antihypertensive medications immediately.
B) Recommend lifestyle modifications and re-evaluate in 3-6 months.
C) Prescribe a combination of ACE inhibitors and beta-blockers.
D) Refer the patient to a cardiologist for further management.

Answer: B

Q9
Question: A 55-year-old male with newly diagnosed stage 1 hypertension (blood pressure 138/88 mmHg) and no evidence of target organ damage or cardiovascular disease presents for counseling. According to the latest guidelines, what is the best initial management approach for this patient?
A) Initiate an ACE inhibitor
B) Start a thiazide diuretic
C) Recommend lifestyle modifications only
D) Refer to cardiology for further evaluation

Answer: C

Q10
Question: A 55-year-old male presents for a routine visit. His blood pressure is 152/94 mmHg measured on two separate occasions. He has no history of cardiovascular disease, diabetes, or renal disease. He adheres to a low-sodium diet and engages in regular aerobic exercise. According to current guidelines, what is the next best step in managing this patient's hypertension?
A) Continue lifestyle modifications and recheck BP in 6 months
B) Initiate low-dose aspirin therapy and recheck BP in 1 month
C) Prescribe a thiazide diuretic and schedule follow-up in 4 weeks
D) Start an ACE inhibitor and monitor kidney function

Answer: C

## Managing the patient with emotional distress (EPA)

Q1
Question: A 45-year-old man presents to your office expressing increased emotional distress and difficulty sleeping after losing his job one month ago. He denies suicidal thoughts but feels overwhelmed and persistently sad. What is the most appropriate initial management step?
A) Prescribe an SSRI to address depressive symptoms
B) Recommend cognitive behavioral therapy (CBT)
C) Provide supportive counseling and assess for additional stressors
D) Refer immediately to psychiatry for specialized care

Answer: C

Q2
Question: A 45-year-old female presents to your clinic with complaints of increased irritability, insomnia, and difficulty concentrating for the past month. She recently lost her job and reports feeling "overwhelmed." She denies suicidal ideation but admits to feeling hopeless at times. Which of the following is the most appropriate initial step in managing her emotional distress?
A) Prescribe a selective serotonin reuptake inhibitor (SSRI) immediately
B) Recommend regular physical exercise and mindfulness techniques
C) Refer to a psychiatrist for evaluation of major depressive disorder
D) Provide supportive communication and assess for social support

Answer: D

Q3
Question: A 45-year-old male presents with complaints of fatigue, difficulty concentrating, changes in appetite, and feelings of hopelessness for the past six weeks. You suspect major depressive disorder. What is the most appropriate initial step in managing this patient's emotional distress?
A) Initiate psychotherapy and refer to a psychiatrist for further evaluation
B) Prescribe a selective serotonin reuptake inhibitor (SSRI) immediately and schedule a follow-up in one week
C) Perform a detailed risk assessment for suicidality and assess for comorbid conditions
D) Recommend regular physical exercise and reassessment in one month

Answer: C

Q4
Question: A 45-year-old woman presents with symptoms of emotional distress related to her recent job loss and reports difficulty sleeping, poor appetite, and frequent crying. Which of the following initial interventions is most appropriate to address her symptoms and support her mental well-being?
A) Prescribe a selective serotonin reuptake inhibitor (SSRI).
B) Refer her for immediate inpatient psychiatric care.
C) Provide supportive counseling and explore coping strategies.
D) Recommend over-the-counter melatonin for sleep difficulties.

Answer: C

Q5
Question: A 45-year-old female presents to your clinic with complaints of fatigue, difficulty concentrating, and frequent episodes of crying over the past two months. She lost her job recently and reports feelings of worthlessness and hopelessness but denies suicidal ideation. What is the most appropriate initial step in managing this patient's emotional distress?
A) Prescribe an SSRI and schedule a follow-up visit in one month
B) Refer the patient to a psychiatrist for evaluation and therapy

C) Conduct a thorough evaluation for signs of major depressive disorder and provide supportive counseling
D) Recommend aerobic exercise and mindfulness meditation as the primary intervention

Answer: C

Q6

Question: A 35-year-old female presents to your clinic with complaints of persistent sadness, poor sleep, fatigue, and difficulty concentrating for the past month. She denies suicidal ideation. She reports a recent divorce and increased stress at work. What is the most appropriate initial management for this patient with emotional distress?

A) Prescribe a selective serotonin reuptake inhibitor (SSRI)
B) Recommend cognitive behavioral therapy (CBT) and schedule follow-up
C) Refer to psychiatry for evaluation and management
D) Order laboratory tests to rule out underlying medical causes

Answer: B

Q7
Question: A 45-year-old woman presents to your clinic reporting persistent sadness, frequent crying spells, and difficulty sleeping over the past two months following the death of her spouse. She denies suicidal thoughts but states she struggles to find enjoyment in her usual activities. She has no significant medical history and denies substance use. Which of the following is the most appropriate initial management step?
A) Prescribe a selective serotonin reuptake inhibitor (SSRI) for depression
B) Refer to a psychiatrist for evaluation of major depressive disorder
C) Provide supportive counseling and encourage healthy coping mechanisms
D) Perform a suicide risk assessment and hospitalize if deemed high-risk

Answer: C

Q8
Question: A 45-year-old male presents to the clinic with complaints of insomnia, fatigue, and difficulty concentrating after the recent death of his spouse two months ago. He exhibits a flat affect and reports feeling hopeless but denies suicidal ideation. What is the most appropriate initial intervention?
A) Prescribe a selective serotonin reuptake inhibitor (SSRI) immediately
B) Refer the patient to inpatient psychiatric care
C) Explore the patient's grief and offer supportive counseling
D) Recommend over-the-counter melatonin for insomnia

Answer: C

Q9
Question: A 45-year-old patient presents to your office with complaints of fatigue, poor concentration, and sleep disturbances for six weeks. They recently went through a divorce and report feeling overwhelmed and tearful daily. What is the most appropriate initial management strategy for this patient presenting with emotional distress?
A) Prescribe a selective serotonin reuptake inhibitor (SSRI)
B) Recommend cognitive behavioral therapy (CBT) and provide supportive counseling
C) Order extensive laboratory testing, including thyroid function tests and serum cortisol levels
D) Refer to psychiatry for immediate evaluation

Answer: B

Q10
Question: A 45-year-old woman presents to your office with complaints of feeling overwhelmed, irritable, and having difficulty sleeping over the past month due to marital problems and job stress. At times, she feels tearful but denies active suicidal ideation, psychosis, or substance use. What is the most appropriate initial approach to manage her emotional distress?
A) Advise starting a selective serotonin reuptake inhibitor (SSRI) immediately
B) Recommend referral to inpatient psychiatric care
C) Validate her feelings, provide emotional support, and consider referral to counseling
D) Prescribe oral benzodiazepines for short-term symptom control

Answer: C) Validate her feelings, provide emotional support, and consider referral to counseling

**Conducting the well-adult visit (EPA)**

Q1

Question: Which of the following components is considered essential during a well-adult visit to address preventive care in a comprehensive manner?
A) Assessment of occupational stress only
B) Recommendation for routine imaging studies
C) Review of immunization history and administration of vaccines as indicated
D) Prescription of multivitamins for all patients

Answer: C

Q2

Question: During a well-adult visit, which of the following is the most appropriate initial step in addressing a patient's concern about occasional chest discomfort?
A) Order an electrocardiogram (ECG) immediately.
B) Perform a detailed review of the patient's risk factors for cardiovascular disease.
C) Refer the patient to a cardiologist for further evaluation.
D) Prescribe a proton pump inhibitor for potential gastroesophageal reflux disease.

Answer: B

Q3

Question: During a well-adult visit, which of the following is the most appropriate initial screening test for detecting alcohol use disorder?
A) Liver function tests
B) Serum carbohydrate-deficient transferrin (CDT)
C) AUDIT-C questionnaire
D) Genetic testing for alcohol predisposition

Answer: C

Q4

Question: During a well-adult visit for a 45-year-old male with no significant medical history, which of the following screenings is recommended according to USPSTF guidelines?

A) Annual electrocardiogram
B) Prostate-specific antigen (PSA) screening
C) Colorectal cancer screening
D) Hepatitis C screening

Answer: C

Q5
Question: A 50-year-old male presents for a well-adult visit. He has no significant complaints and no chronic medical conditions. According to current guidelines, which screening should be offered?
A) Prostate-specific antigen (PSA) screening
B) Colonoscopy or other colorectal cancer screening
C) Electrocardiogram (ECG)
D) Serum testosterone measurement

Answer: B

Q6
Question: Which of the following is the most appropriate initial step in conducting a well-adult visit?
A) Order extensive laboratory testing based on age-related guidelines
B) Perform a comprehensive physical examination without addressing patient concerns
C) Establish an agenda collaboratively with the patient to address health goals and priorities
D) Focus primarily on discussions about specialist referrals

Answer: C

Q7
Question: During a well-adult visit, which screening test is recommended for all adults aged 50 to 75 years to detect colorectal cancer, according to current USPSTF guidelines?
A) Prostate-specific antigen (PSA) test
B) Fecal occult blood testing (FOBT)
C) Hepatitis C antibody screening

D) Low-dose computed tomography (LDCT)

Answer: B

Q8
Question: During a well-adult visit, which preventive measure is recommended for an average-risk 50-year-old patient who has no family history of colorectal cancer?
A) Colonoscopy every year starting at age 50
B) Fecal immunochemical test (FIT) annually starting at age 50
C) Colonoscopy every 10 years starting at age 45
D) Flexible sigmoidoscopy every 5 years starting at age 40

Answer: B

Q9
Question: A 45-year-old male presents for his annual well-adult visit. He is asymptomatic and has no significant medical history. According to current USPSTF recommendations, which of the following screening tests should be routinely offered at this visit?
A) Serum prostate-specific antigen (PSA)
B) Colonoscopy for colorectal cancer screening
C) Lipid panel for dyslipidemia screening
D) Chest X-ray for lung cancer screening

Answer: C

Q10
Question: During a well-adult visit, which of the following best reflects the recommended frequency for screening for alcohol misuse in asymptomatic adults according to USPSTF guidelines?
A) Screening should be conducted every 5 years.
B) Screening should be conducted annually.
C) Screening should be conducted only if the patient reports relevant symptoms.
D) Screening should be conducted at least once in adulthood, with subsequent screenings determined by clinical judgment.

Answer: B

# Guided Approach Questions

**Acute Care and Diagnosis > Respiratory > Flu and Other Viral Pneumonia (10 questions)**

Q1
Question: A 32-year-old female presents with cough, fever, myalgias, and malaise for the past 5 days. She reports recent exposure to a family member diagnosed with influenza. On examination, she has a temperature of 101.3°F, respiratory rate of 18/min, and oxygen saturation of 96% on room air. Chest auscultation reveals scattered rhonchi. Which of the following is the most appropriate initial diagnostic step?

A) Chest X-ray
B) Rapid influenza diagnostic test
C) Sputum culture
D) Polymerase chain reaction (PCR) for respiratory pathogens

Answer: B

Q2
Question: A 44-year-old female presents to the clinic with fever, fatigue, nonproductive cough, and dyspnea. A chest radiograph reveals diffuse bilateral infiltrates. Nasopharyngeal swab confirms influenza A virus. Which one of the following is the most appropriate initial management?
A) Oral azithromycin
B) Supportive care and oral oseltamivir
C) Intravenous ceftriaxone and vancomycin
D) Oral corticosteroids

Answer: B

Q3
Question: A 65-year-old male with a history of COPD presents to the emergency department with fever, a productive cough, and shortness of breath for the past 3 days. On examination, he appears ill, his temperature is 39.2°C (102.6°F), pulse is 110/min, respiratory rate is 28 breaths/min, and oxygen saturation is 91% on room air. Chest auscultation reveals diffuse crackles. A chest X-ray demonstrates bilateral interstitial infiltrates. Influenza is suspected, and nasopharyngeal swab results confirm influenza A. What is the most appropriate next step in management?

A) Initiate corticosteroids and broad-spectrum antibiotics
B) Administer oseltamivir and supportive care
C) Initiate bronchodilators and oxygen therapy only
D) Perform sputum culture and await results before treatment

Answer: B) Administer oseltamivir and supportive care

Q4
Question: A 42-year-old male presents to the clinic with fever, dyspnea, and a nonproductive cough lasting 5 days. He has a history of asthma but is otherwise healthy. Despite supportive care, his symptoms worsen. Chest radiography reveals bilateral patchy infiltrates. Influenza A is confirmed via PCR from a nasopharyngeal swab. Which of the following is the most appropriate next step in management?

A) Supportive care only
B) Initiation of oseltamivir
C) Empiric broad-spectrum antibiotics
D) Pulmonary function testing

Answer: B

Q5
Question: A previously healthy 46-year-old male presents to the emergency department with 3 days of fever, nonproductive cough, and progressive dyspnea. Physical examination reveals tachypnea, crackles bilaterally, and hypoxia with an oxygen saturation of 88% on room air. A chest radiograph shows diffuse interstitial infiltrates. Polymerase chain reaction testing of a nasopharyngeal swab confirms influenza A infection. Which one of the following is the most appropriate initial management for this patient?

A) Amoxicillin-clavulanate
B) Oseltamivir
C) Corticosteroids
D) Azithromycin

Answer: B

Q6
Question: A 65-year-old female presents with fever, dyspnea, and a severe nonproductive cough for the past five days. She reports recent contact with her grandson, who had influenza symptoms last week. Chest auscultation reveals diffuse crackles, but no egophony or focal consolidation. Chest radiography shows bilateral interstitial infiltrates. Based on this presentation, which of the following pathogens is the most likely cause of her condition?

A) Streptococcus pneumoniae
B) Influenza virus
C) Mycoplasma pneumoniae
D) Aspergillus fumigatus

Answer: B

Q7
Question: A 34-year-old female presents with fever, cough, and progressive shortness of breath for the past week. She reports flu-like symptoms prior to the onset of respiratory complaints. Chest radiography reveals diffuse bilateral infiltrates. Which of the following is the most likely diagnosis?

A) Bacterial pneumonia
B) Influenza-associated viral pneumonia
C) Mycobacterium tuberculosis infection
D) Pulmonary embolism

Answer: B

Q8
Question: A 42-year-old male presents with fever, nonproductive cough, and progressive shortness of breath for the past week. Chest radiography reveals diffuse interstitial infiltrates. The patient has a history of recent travel during the peak influenza season and denies exposures to sick contacts. Rapid influenza testing is positive. Which of the following is the most appropriate management for suspected influenza-associated viral pneumonia?

A) Administer oral oseltamivir
B) Initiate broad-spectrum antibiotics
C) Perform bronchoscopy for evaluation

D) Begin corticosteroids

Answer: A

Q9
Question: A 72-year-old male presents to the emergency department with fever, cough, and progressive shortness of breath over the last 6 days. His symptoms began after attending a family gathering during flu season. Chest auscultation reveals bilateral crackles, and a chest X-ray shows multifocal infiltrates. Which of the following is the most likely causative pathogen for his symptoms given the seasonal context and clinical presentation?

A) Streptococcus pneumoniae
B) Influenza virus
C) Mycoplasma pneumoniae
D) Legionella pneumophila

Answer: B

Q10
Question: A 67-year-old male with a history of COPD and heart failure presents to the clinic with fever, dyspnea, and a nonproductive cough for 3 days. He reports feeling lethargic and notes mild confusion. His temperature is 101.5°F, respiratory rate is 24 breaths per minute, and oxygen saturation is 88% on room air. Chest radiography reveals bilateral infiltrates. Laboratory studies are notable for lymphocytosis. Which one of the following pathogens is the most likely cause of his symptoms?

A) Streptococcus pneumoniae
B) Influenza virus
C) Legionella pneumophila
D) Respiratory syncytial virus (RSV)

Answer: B

**Chronic Care Management > Cardiovascular > Hypertension (10 questions)**

Q1
Question: A 48-year-old male with a history of hypertension presents for follow-up. His current medications include lisinopril and hydrochlorothiazide. His blood pressure averages 152/94 mm Hg during serial office visits despite adherence to therapy and lifestyle modifications. Which one of the following is the next best step in managing his hypertension?

A) Increase the dose of lisinopril
B) Replace hydrochlorothiazide with amlodipine
C) Add spironolactone to his regimen
D) Perform ambulatory blood pressure monitoring

Answer: C

Q2
Question: A 48-year-old male with a history of hypertension presents for routine follow-up. His blood pressure readings at home have been consistently elevated despite adherence to a low-sodium diet and regular exercise. He is currently taking lisinopril. On examination, his blood pressure is 158/92 mm Hg in both arms. Laboratory studies are within normal limits, including renal function and serum potassium. Which one of the following is the most appropriate next step in management?
A) Add hydrochlorothiazide.
B) Increase the dosage of lisinopril.
C) Switch lisinopril to amlodipine.
D) Refer for renal artery imaging.

Answer: A) Add hydrochlorothiazide.

Q3
Question: A 52-year-old male with a history of hypertension presents for follow-up. His blood pressure readings average 152/94 mm Hg despite adherence to lifestyle modifications and treatment with hydrochlorothiazide. His BMI is 30 kg/m², and labs show normal renal function. Which one of the following is the most appropriate next step in managing his hypertension?

A) Add amlodipine
B) Add lisinopril
C) Increase hydrochlorothiazide dosage
D) Switch to a beta-blocker

Answer: B

Q4
Question: A 65-year-old male with a history of hypertension and diabetes mellitus visits your clinic for follow-up regarding his blood pressure management. His current blood pressure readings average 152/94 mm Hg, despite adherence to a low-sodium diet and regular exercise. He is taking lisinopril (Zestril) 20 mg daily and his renal function and potassium levels are within normal limits. Which of the following is the most appropriate next step in his management?

A) Increase the dose of lisinopril (Zestril)
B) Add a thiazide diuretic, such as hydrochlorothiazide
C) Add a beta blocker, such as metoprolol
D) Refer for renal artery imaging to rule out secondary hypertension

Answer: B

Q5
Question: A 55-year-old male with a history of poorly controlled hypertension presents with sudden-onset severe headache and nausea. His blood pressure is measured at 210/120 mmHg. Physical examination reveals poorly reactive pupils and focal weakness on his right side. Which of the following is the most likely diagnosis?
A) Hypertensive urgency
B) Hypertensive emergency
C) Primary intracranial tumor
D) Subacute thyroiditis

Answer: B

Q6
Question: A 60-year-old male with a history of type 2 diabetes mellitus and hyperlipidemia presents for routine follow-up. His blood pressure readings over the last three visits have averaged 145/95 mm Hg despite lifestyle modifications. According to current guidelines, which of the following is the most appropriate initial pharmacologic therapy for his hypertension?
A) Hydrochlorothiazide
B) Lisinopril
C) Amlodipine
D) Metoprolol

Answer: B) Lisinopril

Q7
Question: A 48-year-old male with a history of obesity and type 2 diabetes presents to your clinic for chronic care management. His blood pressure is consistently elevated, with readings averaging 152/90 mmHg over several visits. Laboratory results are notable for normal renal function. Lifestyle modifications have been attempted but have yielded minimal improvement. According to current guidelines, which one of the following is the most appropriate initial pharmacologic therapy for this patient?

A) Thiazide diuretics
B) Angiotensin-converting enzyme (ACE) inhibitors
C) Calcium channel blockers
D) Beta-blockers

Answer: B) Angiotensin-converting enzyme (ACE) inhibitors

Q8
Question: A 45-year-old male with a history of primary hypertension presents for a follow-up visit. His blood pressure is consistently elevated despite adherence to lifestyle modifications, including a low-sodium diet and exercise. He is currently taking lisinopril (Zestril). Recent lab results show a serum potassium level of 4.2 mEq/L (N: 3.5–5.1) and a serum creatinine level of 1.0 mg/dL (N: 0.6–1.1). You decide to add a medication to his therapy. Which one of the following is the most appropriate next step in his management?

A) Hydrochlorothiazide
B) Amlodipine (Norvasc)
C) Spironolactone (Aldactone)
D) Metoprolol (Lopressor)

Answer: A) Hydrochlorothiazide

Q9
Question: A 55-year-old male with a history of obesity, type 2 diabetes mellitus, and chronic hypertension presents for follow-up. His blood pressure readings over the last month have been consistently elevated, averaging 158/96 mm Hg despite adherence to his lifestyle changes and medication regimen. He is currently taking amlodipine and lisinopril. What is the most appropriate next step in the management of his hypertension?

A) Increase the dose of lisinopril
B) Add hydrochlorothiazide to the treatment regimen
C) Substitute amlodipine with metoprolol
D) Refer the patient for secondary hypertension workup

Answer: B

Q10
Question: A 60-year-old male with a longstanding history of hypertension presents for routine evaluation. Despite consistent adherence to a prescribed lifestyle modification plan, his blood pressure remains significantly elevated at 160/100 mm Hg. Which one of the following is the most appropriate next step in management?
A) Begin low-dose aspirin therapy
B) Initiate a calcium channel blocker
C) Recommend re-evaluation in six months
D) Perform ambulatory blood pressure monitoring

Answer: B) Initiate a calcium channel blocker

**Emergent and Urgent Care > Psychiatric/Behavioral > Mental Health Emergencies (10 questions)**

Q1
Question: A 32-year-old male with a history of schizophrenia is brought to the emergency department by his family after threatening self-harm and experiencing auditory hallucinations. He is agitated, stating that voices are telling him to hurt himself. What is the most appropriate initial step in managing this patient?
A) Begin treatment with a selective serotonin reuptake inhibitor (SSRI)
B) Administer haloperidol and monitor for response
C) Provide verbal de-escalation and encourage outpatient follow-up
D) Evaluate medical causes and initiate sedation for acute agitation

Answer: D

Q2
Question: A 45-year-old male with a history of bipolar disorder presents to the emergency department with severe agitation, pressured speech, and paranoia. He reports not sleeping or eating for three days and states, "I have a big project that will change the world." Which one of the following is the most appropriate initial intervention?
A) Administer oral lorazepam
B) Initiate lithium therapy
C) Provide a low-stimulation environment
D) Administer intramuscular haloperidol

Answer: D

Q3
Question: A 22-year-old male presents to the emergency department with sudden-onset auditory hallucinations. He is pacing and agitated, starts yelling incoherently, and attempts to strike a nurse. His medical record shows no history of psychiatric illness or substance use. What is the most appropriate initial management for this patient's acute agitation?

A) Verbal de-escalation strategies alone
B) Restraints without medication
C) Intramuscular haloperidol and lorazepam
D) Initiation of long-term antipsychotic medication

Answer: C

Q4
Question: A 28-year-old male is brought to the emergency department by law enforcement after being found wandering along a highway. He appears highly agitated, has disorganized speech, and is repeatedly shouting that cars are trying to communicate with him. He refuses to provide a full history and resists physical examination but has no outward signs of trauma. What is the most appropriate initial management in this situation?

A) Administer benzodiazepines for agitation
B) Initiate antipsychotic treatment to address psychosis
C) Attempt verbal de-escalation and build rapport
D) Use physical restraints immediately to prevent harm to self or others

Answer: C

Q5
Question: A 28-year-old male with a history of schizophrenia presents to the emergency department with aggressive behavior and acute agitation. He is disoriented and is not responding to verbal cues. Physical examination shows elevated blood pressure and heart rate, and he does not seem to be withdrawing from any substances. After ensuring staff safety, what is the most appropriate initial management for this patient?

A) Start oral second-generation antipsychotics immediately
B) Administer intramuscular benzodiazepines
C) Begin supportive care and observation only

D) Administer intramuscular haloperidol with or without lorazepam

Answer: **D**

## Q6

Question: A 32-year-old male is rushed to the emergency department by family members due to sudden agitation, psychosis, and paranoia. He has a history of schizophrenia and is non-adherent to his antipsychotic medications. He is pacing the room, shouting, and threatening others. What is the most appropriate initial management for this patient in this acute psychotic episode?

A) Administer lorazepam intramuscularly for sedation
B) Initiate risperidone orally for psychosis control
C) Utilize physical restraints and observe without pharmacologic intervention
D) Administer haloperidol intramuscularly for immediate symptom control

Answer: D

## Q7

Question: A 28-year-old male presents to the emergency department after expressing intent to harm himself. He reports recent loss of employment and relationship difficulties. He states he has access to a firearm at home. Which one of the following is the most appropriate immediate action?

A) Prescribe an antidepressant and schedule outpatient therapy
B) Provide verbal reassurance and discharge with instructions to return if symptoms worsen
C) Initiate psychiatric hospitalization for suicide risk management
D) Refer to a community support group for coping strategies

Answer: C

## Q8

Question: A 32-year-old male presents to the emergency department with agitation, delusions, and violent behavior. He has no prior psychiatric history but admits to intermittent use of methamphetamine. On examination, his vital signs are: blood pressure 180/100 mm Hg, heart rate 120 bpm, and temperature 101.2°F (38.4°C). His drug screening is positive for methamphetamine. The most appropriate initial management of this patient includes:
A) Physical restraints alone
B) Haloperidol with close cardiac monitoring
C) Lorazepam to manage agitation and hyperthermia

D) Immediate psychiatric hospitalization

Answer: C) Lorazepam to manage agitation and hyperthermia

Q9
Question: A 28-year-old male is brought to the emergency department by EMS following an intentional overdose of prescription medications. He is agitated and non-cooperative during evaluation. His vital signs include a heart rate of 115 bpm, respiratory rate of 22/min, and blood pressure of 150/95 mm Hg. An ECG shows sinus tachycardia without ischemic changes. He has no known psychiatric history, and his family reports recent episodes of insomnia and increased irritability. What is the most appropriate initial management in this acute psychiatric emergency?
A) Begin psychotherapy sessions immediately
B) Administer intramuscular benzodiazepines
C) Initiate gastric lavage
D) Proceed with physical restraint only

Answer: B) Administer intramuscular benzodiazepines

Q10
Question: A 28-year-old male is brought to the emergency department due to aggressive behavior and agitation. He has a history of schizophrenia and poor medication adherence. On arrival, vital signs are normal, but he is pacing and shouting. Attempts to verbally de-escalate the situation are unsuccessful, and he is at risk of harming himself or others. What is the most appropriate initial pharmacologic treatment in this mental health emergency?
A) Oral haloperidol
B) Intramuscular olanzapine
C) Oral clonazepam
D) Intravenous diazepam

Answer: B

**Preventive Care > Immunization > Immunization Administration and Counseling (10 questions)**

Q1
Question: A 4-year-old child presents to your clinic for a well-child visit. The parent expresses concerns about vaccine safety and requests alternative vaccination schedules. How should you counsel the parent regarding vaccine administration?
A) Agree to the parent's requested schedule to maintain trust.

B) Stress that delaying vaccines increases the risk of preventable diseases.
C) Offer to reduce the number of vaccines given per visit to address concerns.
D) Inform the parent that alternative schedules are safer and endorsed.

Answer: B

Q2
Question: A mother brings her 6-month-old infant to your clinic for routine vaccinations. She expresses concern that receiving multiple vaccines during a single visit may overwhelm the infant's immune system. Which of the following is the most appropriate response to address her concern?

A) Spreading out vaccines over multiple visits reduces the risk of overwhelming the immune system.
B) Receiving multiple vaccines at once is safe and helps protect against serious diseases earlier.
C) Vaccination should be delayed until the infant's immune system is more mature.
D) Administering fewer vaccines per visit reduces the risk of vaccine-associated side effects.

Answer: B

Q3
Question: A 4-month-old infant visits your clinic for a routine wellness check. The parent expresses concern about vaccine safety and hesitates to vaccinate the child. Which of the following is the most appropriate response to address the parent's concerns?
A) "Vaccines are mandatory by law, so you must proceed with vaccination."
B) "Vaccines can occasionally cause severe side effects, so be cautious when deciding."
C) "Vaccines are extensively studied and proven to be safe, protecting children from serious illnesses."
D) "You should read online forums and decide based on your comfort level."

Answer: C

Q4
Question: A 12-month-old child presents for a routine well-child visit. The parent expresses concerns regarding pain and potential side effects from vaccines. You explain the importance of immunizations and address their concerns. Which of the following strategies is most effective in improving vaccine acceptance during counseling?

A) Emphasize societal benefits of vaccination.

B) Focus primarily on the rare risk of adverse effects.
C) Provide clear, evidence-based information about vaccine safety and benefits.
D) Allow the parent to defer all vaccinations until they feel comfortable.

Answer: C

Q5
Question: A 32-year-old patient is nervous about receiving the influenza vaccine because they believe it might cause influenza. What is the most appropriate counseling response?

A) Explain that the vaccine contains live influenza virus and could cause a mild case of influenza.
B) Reassure the patient that the vaccine contains inactivated virus and cannot cause influenza, but mild side effects like soreness or fever are possible.
C) Recommend postponing the vaccine due to patient fear of adverse effects.
D) Advise the patient to take antibiotics in case of side effects after vaccination.

Answer: B

Q6
Question: A 6-month-old infant is brought in for routine well-child care. The parent expresses concern about vaccine safety and hesitates to proceed with the recommended vaccinations. Which of the following is the most appropriate initial response to address the parent's concerns?

A) Provide factual information about vaccine safety and efficacy while listening to the parent's concerns.
B) Advise the parent that delaying vaccines may result in permanent exclusion from daycare.
C) Reassure the parent that vaccine-preventable diseases are now very rare, making vaccination less urgent.
D) Recommend postponing vaccination until the parent feels more comfortable.

Answer: A

Q7
**Question:** A 35-year-old male with no significant past medical history presents to your clinic for a routine wellness visit. He works in construction, travels frequently for work (including international destinations), and has not received any vaccinations in over 15 years. Which one of the following vaccines is most appropriate to administer during today's visit?
A) Human papillomavirus (HPV) vaccine

B) Measles, mumps, and rubella (MMR) vaccine
C) Tetanus, diphtheria, and acellular pertussis (Tdap) vaccine
D) Pneumococcal polysaccharide vaccine

**Answer:** C

Q8
Question: A parent brings their 6-month-old infant to your clinic for a routine check-up. The child has no prior vaccinations. The parent expresses concern about vaccine safety and requests information on which vaccines are recommended at this age. Based on the CDC immunization schedule, which one of the following vaccines is recommended for administration at this visit?

A) MMR (measles, mumps, rubella) vaccine
B) Varicella (chickenpox) vaccine
C) Hepatitis B vaccine (third dose)
D) HPV (human papillomavirus) vaccine

Answer: C) Hepatitis B vaccine (third dose)

Q9
Question: A 25-year-old female presents for routine preventive care and is found to be due for her tetanus immunization. During counseling, she mentions she is considering pregnancy in the next few months and is concerned about vaccine safety. What is the most appropriate recommendation regarding tetanus vaccination for this patient?

A) Postpone tetanus vaccination until after pregnancy
B) Administer Tetanus, Diphtheria, and Acellular Pertussis (Tdap) vaccine now
C) Administer only tetanus and diphtheria (Td) vaccine to minimize risks
D) Avoid all vaccinations during pregnancy

Answer: B) Administer Tetanus, Diphtheria, and Acellular Pertussis (Tdap) vaccine now

Q10
Question: A 2-month-old infant presents for a routine well-child visit. The parents express concern about the number of vaccinations given at this age and worry that they may "overload the immune system." How should the physician counsel the parents regarding the safety and efficacy of routine vaccinations?

A) Postpone administering vaccines until the child is older to reduce immune system stress

B) Recommend vaccinating for only the most critical diseases at this visit to limit exposure
C) Explain that an infant's immune system can handle multiple vaccines without being "overloaded" because they are exposed to more antigens in daily life
D) Suggest spacing out the vaccines over several visits to ensure better tolerance

Answer: C